\documentclass{vgtc}                          % final (conference style)

\graphicspath{{figures/}{pictures/}{images/}{./}} % where to search for the images

\usepackage{times}                     % we use Times as the main font
\usepackage{comment}
\usepackage{cite}
\usepackage{enumitem}
\usepackage{tabu}                      % only used for the table example
\usepackage{booktabs}                  % only used for the table example
\usepackage{lipsum}                    % used to generate placeholder text
\usepackage{mwe}                       % used to generate placeholder figures

\usepackage{mathptmx}                  % use matching math font
\usepackage{mathptmx}                  % use matching math font
\usepackage{amsmath}                   % for align environment
\usepackage{amsmath, amssymb}

\onlineid{0}

\vgtccategory{Research}

\vgtcinsertpkg

\title{Scale-Aware Navigation of Astronomical Survey Imagery Data on High Resolution Immersive Displays}
\author{
Ava Nederlander\thanks{e-mail: ava.nederlander@stonybrook.edu}\\
\parbox{1.4in}{\scriptsize \centering Stony Brook University}
\and
Zainab Aamir\thanks{e-mail: zaamir@cs.stonybrook.edu}\\
\parbox{1.4in}{\scriptsize \centering Stony Brook University}
\and
Arie E. Kaufman\thanks{e-mail: ari@cs.stonybrook.edu}\\
\parbox{1.4in}{\scriptsize \centering Stony Brook University}
}
\teaser{
  \centering
  \includegraphics[
    width=\linewidth,
    trim=0 600 0 600,
    clip
  ]{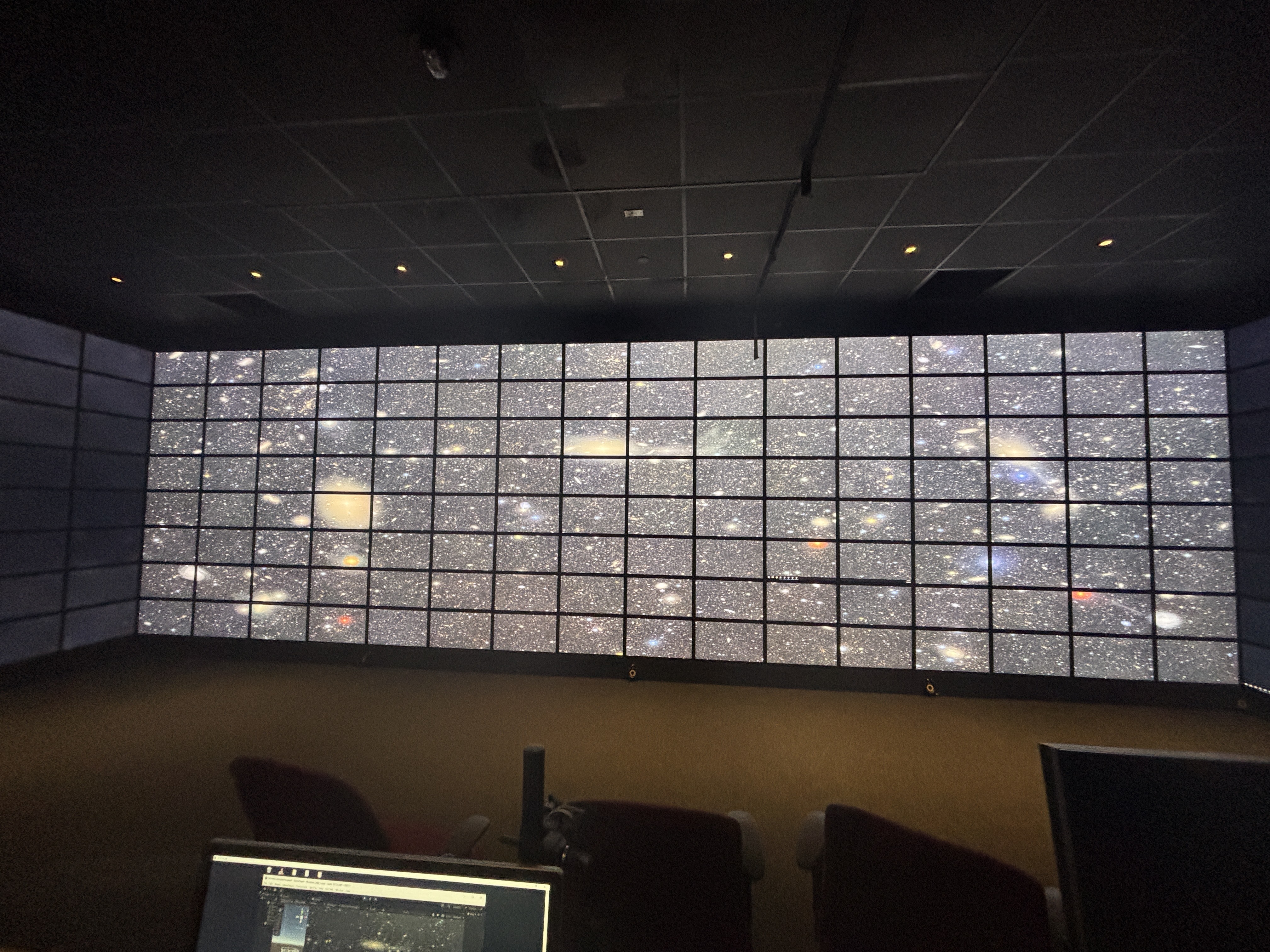}
  \caption{A high-resolution public image from the Vera C. Rubin Observatory displayed on a single wall of the Reality Deck. }
  \label{fig:teaser}
}

\abstract{
Upcoming astronomical surveys produce imagery that spans many orders of magnitude in spatial scale, requiring scientists to reason fluidly between global structure and local detail. Data from the Vera C. Rubin Observatory exemplifies this challenge, as traditional desktop-based workflows often rely on discrete views or static cutouts that fragment context during exploration.

This paper presents a design-oriented framework for scale-aware navigation of astronomical survey imagery in high-resolution immersive display environments. 

We illustrate these principles through representative usage scenarios using Vera Rubin Observatory and Milky Way survey imagery deployed in room-scale immersive environments, including tiled high-resolution displays and curved immersive systems. Our goal is to contribute design insights that inform the development of immersive interaction paradigms for exploratory analysis of extreme-scale scientific imagery.

} 


\begin{document}


\firstsection{Introduction}
\maketitle

Modern astronomical surveys increasingly produce data at scales that challenge conventional visualization and analysis paradigms. The Vera C. Rubin Observatory, for example, is designed around a 3.2-gigapixel camera that repeatedly images the entire visible sky, ultimately producing multi-epoch sky maps containing tens of terapixels of image data over the course of the survey \cite{plazas2025instrument}. While these data are rich in structure, their extreme size and inherent multi-scale nature make it difficult to preserve contextual awareness when examining localized detail using traditional desktop-based visualization tools.

A common strategy for working with large astronomical datasets is to rely on discrete views, static cutouts, or predefined zoom levels \cite{shneiderman1996eyes}. Although effective for focused inspection, these approaches often fragment the exploratory process and obscure relationships between global structure and local phenomena. As a result, scientists are frequently required to mentally reconstruct context when transitioning between scales, which can hinder interpretation, comparison, and collaborative discussion.

Large-scale immersive visualization has previously supported scientific insight in astronomy. Work using full-dome visualization environments, such as those at the Hayden Planetarium, has demonstrated how immersive, high-resolution representations of astronomical data can reveal structural relationships and anomalies that are difficult to perceive using conventional desktop displays \cite{emmart2006digital,emmart2010dome}. These experiences suggest that display scale and immersion can play a meaningful role in exploratory analysis and hypothesis generation, motivating further investigation into immersive environments as tools for scientific sensemaking.

In this work, we focus on interaction design rather than system implementation or empirical evaluation. We articulate a set of design principles for scale-aware navigation of extreme-scale astronomical imagery in high-resolution immersive display environments, centered on continuous zoom interaction, spatial coherence across scale, and layer-based exploration with persistent annotation. These principles are illustrated through representative usage scenarios using Vera C. Rubin Observatory survey data and high-resolution Milky Way imagery \cite{spitzer_milkyway8}, deployed in two complementary immersive display configurations: the Reality Deck, a room-sized installation composed of 416 high-density LCD panels yielding over 1.5 billion pixels across a four-wall surround layout \cite{yost2007realitydeck}, and a large curved immersive environment consisting of 40 high-resolution displays configured to emphasize peripheral visualization and spatial continuity.

By examining how scale-aware navigation principles manifest across these display environments, we highlight how physical extent, aggregate resolution, and embodied interaction shape reasoning across astronomical scales. Our goal is to contribute design insights and practical considerations to ongoing  
%workshop-level 
discussions within the immersive visualization community on interaction paradigms for exploratory analysis of extreme-scale scientific imagery.

\section{Background and Motivation}
\label{sec:background}

\subsection{Astronomical Data at Extreme Scale}

Astronomical survey data is characterized by an extreme range of spatial and temporal scales, requiring analysts to reason across many orders of magnitude within a single dataset \cite{plazas2025instrument,spitzer_glimpse_mipsgal_overview}. Observations may span the full sky while simultaneously supporting inspection of individual sources or localized regions. Scientific tasks often depend on fluid transitions between these scales, such as identifying local anomalies within global distributions or relating small-scale features to large-scale structure.

In common analysis workflows, these scale transitions are mediated through predefined resolutions, separate tools, or static visual products. In this work, we draw on infrared Milky Way imagery from the Spitzer GLIMPSE--MIPSGAL surveys and early public imagery from the Vera C. Rubin Observatory as representative examples of extreme-scale astronomical data used throughout the paper \cite{spitzer_milkyway8,rubin_cosmic_treasure_chest}.

\subsection{Immersive Visualization and High-Resolution Displays}

Immersive visualization environments have been widely studied as platforms for supporting spatial understanding, collaboration, and embodied interaction \cite{Chandler2015ImmersiveA}. High-resolution immersive displays, in particular, combine large physical extent with dense pixel resolution, enabling users to view overview and detail simultaneously without aggressive abstraction or loss of visual fidelity.

Prior work has demonstrated that such environments can support exploratory analysis and collaborative decisionmaking across a variety of domains \cite{ball2007collaborative}. For astronomical survey imagery, where scale is fundamental to interpretation rather than incidental, immersive environments present an opportunity to rethink how users traverse, contextualize, and reason about data across orders of magnitude. This gap motivates our focus on scale-aware navigation principles for immersive exploration of astronomical imagery.

\section{Scale-Aware Immersive Navigation}
\label{sec:immersive}

We frame scale-aware navigation as an interaction paradigm for engaging with extreme-scale astronomical imagery in immersive environments. Rather than treating scale changes as discrete transitions between predefined views, this paradigm emphasizes continuous navigation across orders of magnitude while preserving spatial coherence and orientation. The principles described in this section are inspired by continuous zoom-based navigation systems such as Google Earth, but are adapted to the scale ranges, data characteristics, and analytical goals of astronomical survey imagery.

\subsection{High-Resolution Immersive Display Context}

Our design principles are situated in high-resolution immersive display environments, such as the Reality Deck, which combine large physical extent with dense pixel resolution. These environments allow users to perceive global structure and fine-grained detail simultaneously, without aggressive abstraction or loss of visual fidelity. Physical movement within the space, such as walking, turning, or approaching regions of interest, becomes an integral part of the interaction, complementing traditional input devices and supporting embodied exploration \cite{dourish2001embodied}.

\subsection{Continuous Zoom Navigation Across Scale}

A central principle of scale-aware navigation is continuous zoom interaction across astronomical scales. Users transition smoothly along a single navigation continuum, beginning from full-sky views that reveal large-scale galactic structure, survey coverage, and intensity patterns. As users zoom inward, the visualization progressively reveals finer spatial detail, such as dust lanes, star-forming regions, survey tiles, and individual sources, without breaking contextual continuity.

By preserving spatial relationships throughout navigation, continuous zooming minimizes disruptive context switches and supports reasoning about how local features relate to global structure. While inspired by the interaction paradigm popularized by Google Earth, the intent is not to replicate its interface, but to apply continuous scale transitions to the extreme dynamic ranges characteristic of astronomical survey data.

\subsection{Layered Representations}

Layer-based exploration provides a mechanism for reasoning across scale by allowing users to selectively view and combine multiple data representations, including raw imagery and derived data products. This selective visibility supports tasks such as comparing raw observations to processed results or contextualizing localized features within broader survey context.

\section{Usage Scenarios with Vera Rubin Observatory Data}
\label{sec:scenarios}

To illustrate how scale-aware immersive navigation supports reasoning about extreme-scale astronomical imagery, we describe a set of representative usage scenarios. These scenarios are intended to be illustrative rather than evaluative and are informed by informal demonstrations of the immersive setup to professional astronomers. While these demonstrations helped ground the examples in realistic exploratory practices, they do not constitute a formal user study.

\subsection{From Full-Sky Context to Local Detail}

In this scenario, survey data are presented as a continuous, wrap-around mosaic spanning the walls of the Reality Deck, forming a room-scale full-sky representation. At the overview level, users stand within a panoramic Milky Way projection that encodes large-scale galactic structure, including the disk, bulge, and survey coverage patterns, across the physical extent of the display. The wide field of view allows users to perceive global spatial relationships directly, without the need for panning or repeated reorientation.

Using continuous zoom interaction, users move from this full-sky context toward a specific region of interest while remaining embedded within the surrounding mosaic. As scale increases, aggregated intensity patterns resolve into finer structures such as dust lanes and star-forming regions, and eventually into localized survey tiles and individual sources. Throughout this transition, the surrounding sky remains visible in peripheral vision, preserving awareness of how the selected region is situated within the broader galactic context.

By tightly coupling local inspection with persistent global structure, this form of scale-aware navigation reduces the need for mental reconstruction when transitioning between overview and detail. Users can examine fine-scale features while maintaining spatial continuity across orders of magnitude.

\begin{figure}[t]
  \centering
  \includegraphics[width=\linewidth]{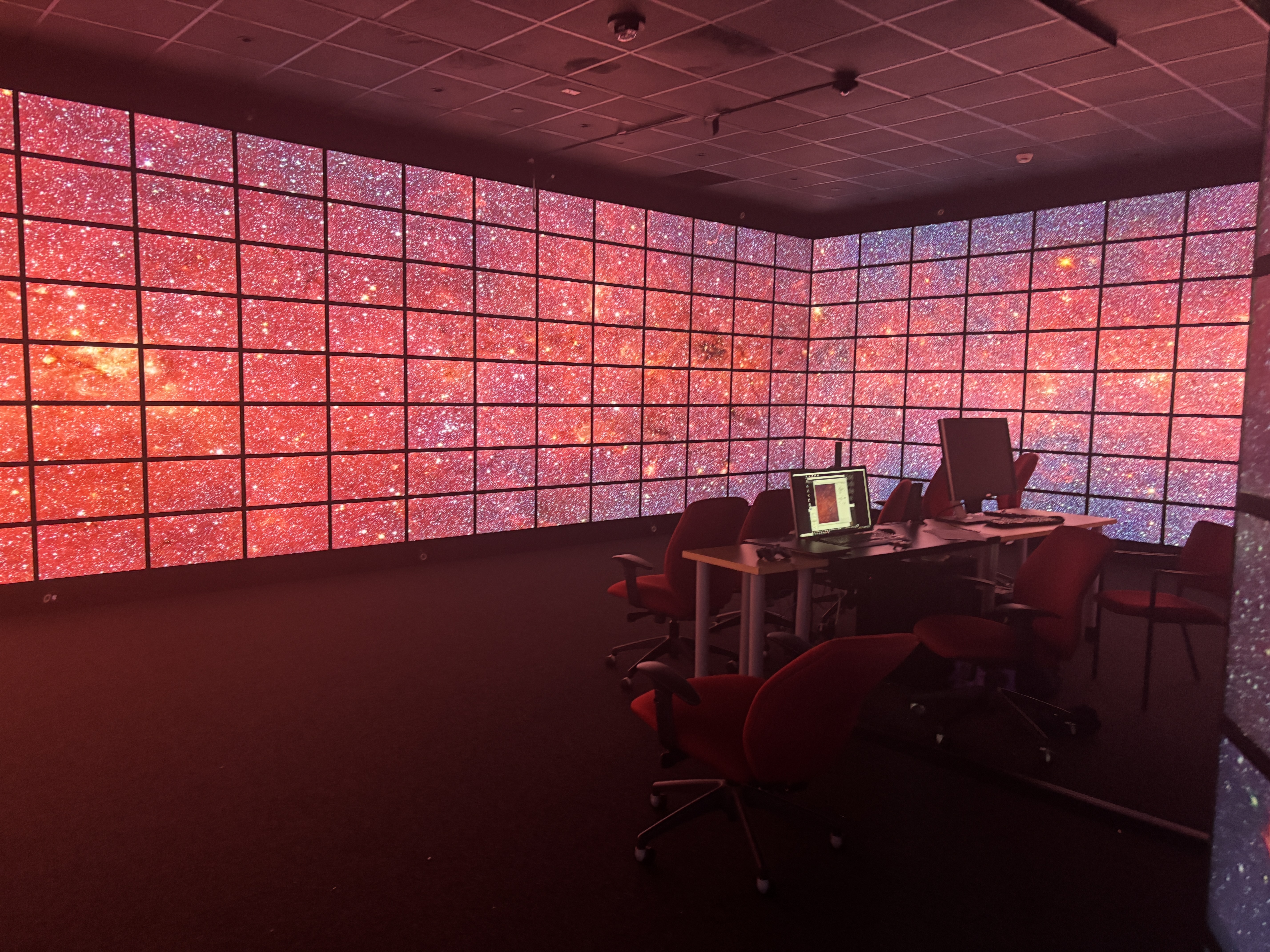}
  \caption{Milky Way infrared imagery from the Spitzer GLIMPSE--MIPSGAL surveys displayed in a high-resolution immersive environment similar to the Reality Deck. The large physical scale and dense pixel resolution support simultaneous perception of global galactic structure and localized detail during continuous zoom-based navigation \cite{spitzer_milkyway8,spitzer_glimpse_mipsgal_overview,yost2007realitydeck}.}
  \label{fig:milkyway_realitydeck}
\end{figure}

Figure~\ref{fig:milkyway_realitydeck} illustrates Milky Way infrared imagery from the Spitzer GLIMPSE--MIPSGAL surveys displayed as a large-scale mosaic in a high-resolution immersive environment. The combination of physical display extent and pixel density enables simultaneous perception of global galactic morphology and localized detail during continuous navigation.

\subsection{Single-Wall and Full-Deck Configuration}

While a full wrap-around configuration emphasizes immersion and spatial orientation, scale-aware navigation can also be supported using a single wall of the immersive display. In this configuration, as shown in Figure~\ref{fig:teaser}, the display functions as a high-resolution focal surface, supporting tasks that require sustained attention to a specific region, detailed comparison between datasets, or small-group discussion centered on a shared view.

In contrast, using all walls of the Reality Deck supports tasks that benefit from peripheral awareness and large-scale spatial reasoning, such as exploring survey coverage, identifying structural relationships across widely separated regions, or collaboratively situating local phenomena within global galactic context. Supporting both configurations allows users to balance immersion, focus, and cognitive load depending on task demands, while retaining the same underlying navigation principles.

\section{Lessons Learned and Design Considerations}
\label{sec:lessons}

Exploring scale-aware navigation of Vera C. Rubin Observatory imagery across different immersive display configurations highlights how interaction design is shaped by display geometry, aggregate resolution, and physical scale. Rather than reporting empirical findings, this section summarizes design considerations intended to inform future immersive visualization work. We organize these observations around two representative classes of immersive environments: high-resolution immersive displays and curved immersive systems.

\subsection{High-Resolution Immersive Displays}

High-resolution immersive display environments combine large physical extent with extremely high aggregate pixel density, enabling users to perceive global structure and fine-grained detail within a single continuous visual field \cite{yost2007realitydeck}. In the context of astronomical imagery, this combination makes continuous navigation across orders of magnitude perceptually meaningful, as fine-scale structures remain visually stable rather than collapsing into abstraction during scale transitions.

The physical scale of these environments foregrounds embodied navigation as a central interaction mechanism. Users naturally move closer to regions of interest, turn to compare distant structures, or step back to regain global context. These bodily actions complement virtual zooming and reinforce spatial understanding, aligning interaction with the physical affordances of the display \cite{dourish2001embodied}. Together, high resolution and physical extent support exploration strategies that integrate overview and detail without relying on discrete view changes.

High-resolution immersive displays are also well suited to shared exploration. Multiple viewers can simultaneously observe and discuss features across scale within a common spatial reference frame, supporting collaborative reasoning without requiring repeated reorientation or turn-taking. These characteristics suggest that such environments are particularly effective for scale-aware navigation paradigms that emphasize continuity and spatial coherence in extreme-scale datasets.

\subsection{Curved Immersive Environments}

Curved immersive environments provide a complementary design space characterized by continuous horizontal fields of view and strong visual immersion. Prior work has shown that curved display geometries can improve spatial awareness and support more natural head- and body-centric navigation compared to flat multi-panel walls \cite{ball2005effects,boorboor2025silo}. For astronomical imagery, this geometry helps maintain contextual continuity across wide visual extents while reducing perceptual discontinuities at display boundaries.

These systems are often paired with head-tracked, viewpoint-corrected rendering, allowing visual content to remain spatially anchored relative to the user. This property supports embodied navigation and spatial comparison through coordinated physical movement and virtual zooming, even when aggregate resolution is lower than that of tiled high-resolution environments. As a result, curved immersive systems can effectively support scale-aware exploration through perceptual continuity and immersion rather than raw pixel density alone.

\subsection{Cross-Environment Design Implications}

Across both display classes, several common design implications emerge. First, preserving continuity across scale is critical regardless of display geometry. Interaction techniques that maintain spatial coherence during navigation help users remain oriented when transitioning across orders of magnitude, while abrupt view changes risk disrupting exploratory reasoning.

Second, embodied interaction consistently plays a central role in how users perceive scale and spatial relationships. Physical movement, head orientation, and body-centered navigation meaningfully contribute to decisionmaking across immersive environments, even when resolution and physical extent differ.

Taken together, these observations suggest that interaction design for extreme-scale astronomical imagery should be grounded in the physical and perceptual affordances of immersive display environments, rather than treating scale navigation as a purely virtual or symbolic operation \cite{Chandler2015ImmersiveA,dourish2001embodied}.

\section{Future Directions}
\label{sec:future}

This work opens several avenues for future research. Formal user studies could examine how scale-aware immersive navigation affects task performance, understanding, and collaboration in exploratory analysis settings \cite{Chandler2015ImmersiveA}. Additional work could investigate how the interaction principles described here generalize to other scientific domains characterized by extreme-scale data. From a systems perspective, future efforts could explore how scale-aware navigation integrates into operational analysis workflows and complements existing desktop-based tools rather than replacing them.

Beyond interaction and systems considerations, immersive high-resolution visualization also presents opportunities for supporting scientific insight and discovery. Prior work in large-scale astronomical visualization, including research conducted in full-dome planetarium environments, suggests that immersive representations can reveal structural relationships and anomalies that may be difficult to perceive using conventional displays \cite{emmart2006digital,emmart2010dome}. Investigating how scale-aware immersive navigation contributes to hypothesis generation and discovery-oriented analysis is an important direction for future study.

Persistent spatial annotation represents a particularly promising extension of the interaction paradigm explored in this work. Anchoring annotations to spatial locations that remain visible across continuous scale changes could support externalization of interpretation, collaborative discussion, and comparative reasoning in shared immersive environments. Future research should examine how annotation design interacts with continuous navigation and layered representations to support coordinated exploratory analysis.

\section{Conclusion}
\label{sec:conclusion}

Extreme-scale astronomical datasets require interaction paradigms that support reasoning across many orders of magnitude. High-resolution immersive displays offer an opportunity to rethink how users navigate, interpret, and discuss such data by preserving spatial continuity across scale. By articulating design principles centered on continuous zoom navigation, spatial coherence, and layered exploration, this work illustrates how immersive environments can support multi-scale reasoning for astronomical imagery. We hope these insights contribute to ongoing 
%workshop-level 
discussions on interaction design for extreme-scale scientific visualization and motivate future investigation into immersive approaches for exploratory analysis and discovery.

\acknowledgments{
 %This research was supported in part by ONR award [insert here] and NSF award [insert here]. Person I acknowledges support from the NSF Graduate Research Fellowship under Grant No. insert here]. %2234683. 
 
 %This research was supported in part by [TBA] award [award number] and [TBA] award [award number]. Person I acknowledges support from the NSF Graduate Research Fellowship under Grant No. [award number]. %2234683. 
 %The authors wish to thank Person II
%Matthew Castellana 
%for advice on Unity and HMD support.
We express appreciation to Simon Birrer and Anja von der Linden for advice about Vera Rubin applications in high-resolution display facilities. 

AN acknowledges support from the NSF Graduate Research Fellowship under Grant No. 2234683. This research was also supported in part by NSF award IIS2529207.

We thank the teams behind the Spitzer Space Telescope GLIMPSE and MIPSGAL surveys for the Milky Way infrared imagery used as an example dataset, and the Vera C. Rubin Observatory / NOIRLab for providing public deep survey imagery that illustrates scale in this work. The authors also acknowledge informal feedback from professional astronomers during demonstrations of the immersive environment. Data credits: Spitzer GLIMPSE–MIPSGAL Milky Way 8 image \cite{spitzer_milkyway8,spitzer_glimpse_mipsgal_overview} and Vera Rubin Observatory imagery \cite{rubin_cosmic_treasure_chest}. }

\bibliographystyle{abbrv-doi}

\bibliography{8references.bib}
\end{document}